# Quantum Hall Measurements of Spatially-Defined, Photoinduced Doping of Graphene/hBN Heterostructures

Son T. Le[1,2*], Thuc T. Mai[3], Maria F. Munoz[3], Angela R. Hight Walker[3*], Curt A. Richter[3], Aubrey T. Hanbicki[2], Adam L. Friedman[2*]

[1] *Institute for Research in Electronics and Applied Physics, University of Maryland, College Park, MD, 20740, USA*
[2] *Laboratory for Physical Sciences, 8050 Greenmead Dr., College Park, MD, 20740, USA*
[3] *National Institute of Standards and Technology, 100 Bureau Dr., Gaithersburg, MD, 20899, USA*

*Corresponding authors: stle@lps.umd.edu, angela.hightwalker@nist.gov, and afriedman@lps.umd.edu

## Abstract

Doped semiconductors are a central and crucial component of all integrated circuits. By using a combination of white light and a focused laser beam, and exploiting *h*BN defect states, heterostructures of *h*BN/Graphene/*h*BN are photodoped *in-operando*, reproducibly and reversibly. We demonstrate device geometries with spatially-defined doping type and magnitude. After each optical doping procedure, magnetotransport measurements including quantum Hall measurements are performed to characterize the device performance. In the unipolar (p+-p-p+ and n-n+-n) configurations, we observe quantization of the longitudinal resistance, proving well-defined doped regions and interfaces that are further analyzed by Landauer-Buttiker modeling. Our unique measurements and modeling of these optically doped devices reveal a complete separation of the p- and n-Landau level edge states. The non-interaction of the edge states results in an observed "insulating" state in devices with a bi-polar p-n-p configuration that is uncommon and has not been measured previously in graphene devices. This insulating state could be utilized in high-performance graphene electrical switches. These quantitative magnetotransport measurements confirm that these doping techniques can be applied to any 2D materials encapsulated within *h*BN layers, enabling versatile, rewritable circuit elements for future computing and memory applications.

**Introduction**

As feature sizes shrink with the continual march toward smaller integrated circuit nodes, accurate spatial doping of semiconductors becomes more challenging.[1] This trend is driven by the need to reduce power consumption and increase speed, while maintaining low cost. The constraints and ultimate limitations of CMOS-based systems have led to the exploration of alternate materials systems and devices heterostructures. Novel two-dimensional (2D) materials such as graphene (Gr),[2] and transition metal dichalcogenides (TMDs),[3] are emerging as possible solutions to achieve computing beyond CMOS.[4] These materials provide new avenues for 3D heterogeneously integrated circuit designs at high densities due to their atomic size and ability to be arbitrarily stacked on top of each other.[5] To date, most 2D material doping techniques are limited to either intrinsic doping or charge transfer. In the former, dopants are incorporated into 2D crystals during synthesis hence modifying the overall doping level.[6] In the latter, electrons are either donated (n-doping) or captured (p-doping) from a material or molecule in direct contact with the top of the 2D material, thus altering the electron/hole density.[7,8] Challenges in both of these approaches include maintaining material quality,[9,10] reproducibly controlling the doping level,[11,12] and intentionally altering the doping once a device has been fabricated.[13] Additionally, doping through adsorption often requires imprecise liquid chemistries that are unlikely to be adopted by the wider semiconductor fabrication industry.[14]

Hexagonal boron nitride (*h*BN) is a superb encapsulation material for graphene and TMDs due to its ability to screen phonon scattering and protect from atmospheric adsorption and oxidation.[15] Devices encapsulated by *h*BN have higher quality electronic[16] and optical[17] performance. Furthermore, *h*BN has naturally occurring defects that were previously considered undesirable, but now are receiving attention for applications as single photon emitters[18] with a nitrogen vacancy or spin-polarized defect states for quantum information processing and sensing.[19,20] Interestingly, a theoretical study proposed that the defects in *h*BN could be used to modulate the carrier density in graphene and other 2D TMDs films.[21] This strategy has the natural benefit of separating the defects from the transport carriers, thus reducing the unwanted impurity scattering and enhancing the mobility and other electronic properties of the active 2D materials. Additionally, in any future high-quality commercial 2D electronic device, *h*BN will likely already be present in the material stack, thus utilizing its defects would not create any additional cumbersome fabrication steps.

Preliminary experimental studies have demonstrated doping in *h*BN/Gr, *h*BN/TMD heterostructures,[22–26] where the doping was realized by optically and/or electrically activating the defect states in *h*BN thereby inducing carrier density changes in the 2D material. These studies focused on uniformly changing the charged defect concentration in the *h*BN substrate to change the overall doping level of the 2D materials (graphene or TMDs) placed on top of the *h*BN. However, some recent work beautifully demonstrated spatial controlled photodoping with defects in *h*BN, [25,26] and ultraviolet (UV) light-activated resist enabling charge transfer. [27,28] Outside the scope of these works, but critically important, is the measurement of the electrostatic doping profile between the spatially doped regions which we present here. Another recent demonstration of spatial doping in an *h*BN/Gr heterostructure[29] employed voltage pulses from a scanning tunneling microscope tip to activate the defect states in *h*BN to remotely dope graphene[30]. This is an elegant way to demonstrate spatial doping and it gives valuable microscopic information about the nature of the doping technique. However, it has limited practical application to large-scale devices.

For practical functional devices, realization and detailed characterization of spatial doping modulation is crucial. In this study, we address this important, unanswered question. The doping level of graphene in an *h*BN/Gr/*h*BN heterostructure is controlled by leveraging a photodoping technique that exploits the defect states in *h*BN that is amenable for large-scale device applications.[21,25] We quantify the tunability, reproducibility, and precision of the doping with quantum Hall magnetotransport measurements. We first modulate the overall doping of a device to a specific level by controlling the back-gate voltage and light exposure. Then, by using a focused laser source, we demonstrate spatially selective doping of the heterostructure device and create engineered regions of modulated charge density. With this method, we demonstrate modification of devices *in-operando* for custom functionalities, such as lateral p-n-p junctions in a way that is amenable to large-scale device fabrication applications. In particular, our quantum Hall transport analysis of the devices verifies the distinct boundaries between the p- and n-doped regions across the entire width of the entire device. The doping uniformity can be constructed *via* the clear observation of "unconventional" Landau levels (LLs) in the longitudinal conductance in the unipolar doping regimes (p+-p-p+ and n-n+-n), which is fully explained within the Landauer-Büttiker Resistance Model [31]. Furthermore, in the p-n-p doping regime, the nature of the interactions of the LL edge states at the interfacial regions between spatially doped areas yields detailed information about the electrostatic doping profile of the system.

The versatility and accuracy of the doping technique when combined with quantum Hall magentotransport measurements that we demonstrate here, has major implications for the development of future memory and computing devices utilizing 2D material heterostructures. Optically activated dopants in $h$BN allows for accurate doping level and also precise spatially-controlled doping. Moreover, these remote dopants help protect the unique electrical, optical, and spintronic properties of 2D materials, in contrast with the unwanted degradation due to carrier scattering and/or material quality reduction due common in other doping processes. Additionally, the reversibility of the method provides a venue to explore unique applications. One possible application is reconfigurable regions of a circuit i.e. field programmable devices, and/or *in situ* error corrections.

## METHODS

### Device Fabrication

The device used in this study was fabricated on a heavily doped $p^+$Si substrate with ~ 290 nm of thermally grown $SiO_2$ on top. The doped Si was used as the backgate. A dry transfer method was used to fabricate and transfer an exfoliated $h$BN/Gr/$h$BN heterostructure onto a substrate with pre-patterned alignment markers.[32] The graphene was a single layer and the thickness of both the top and bottom $h$BN layers was ~ 50 nm. The heterostructure was then patterned into a conventional Hall bar geometry by using electron-beam (e-beam) lithography followed by subsequent reactive ion etching.[32–35] Care was taken to select an area of the heterostructure that was bubble free (as measured by atomic force microscopy (AFM)) for the Hall bar to achieve the highest device quality. A second e-beam lithography step, followed by the e-beam deposition of Cr/Pd/Au (5 nm/10 nm/75 nm) and subsequent metal lift-off, was used to form 1D metal edge contacts to the $h$BN-sandwiched graphene. **Figure 1A** shows the schematic cross-section of the structure with the edge-contacted $h$BN/Gr/$h$BN device at the middle. **Figure 2A** shows a 3D rendered AFM image of the complete device.

The $h$BN used in this study was purchased from 2DSemiconductors.com. [36] It is highly likely that different sources of $h$BN have different defect densities, which play a role in successful doping. A full understanding of the density and nature of these defects is outside the scope of this work as well as is investigated elsewhere. [37]

**Doping Procedure**

All doping results reported in the main text were performed at 1.6 K, the base temperature of the apparatus used to take most of the measurements. However, we found that doping at room temperature was equally as successful (See **Supplemental Information**), as was observed in Ref 26. For the results reported here, we used two general doping procedures: Procedure (1), a global doping, where the entire *h*BN/Gr/*h*BN device was doped at the same level, and Procedure (2), a local doping, where a well-focused 633 nm laser, with ~ 100 uW of laser power, was used to selectively dope only a specific region of the *h*BN/Gr/*h*BN device. The laser wavelength, hence the photon energy used, may affect the type of defect that is activated in the hBN bandgap. Although a different laser was used in Ref 26. Nonetheless, for the n-doping experiment presented here, laser wavelength is not important.

**Procedure (1)**: A fixed backgate voltage, $V_{D-G}$, was applied while the entire *h*BN/Gr/*h*BN heterostructure was illuminated with an LED white-light illuminator focused on the sample through a 50x objective. The illumination time was fixed at 10 minutes for each doping level. After the light source was extinguished, the backgate voltage was set to zero to conclude the doping procedure. **Procedure (1)** is shown schematically in **Figure 1A**.

**Procedure (2)**: The sample is mounted on a magnetic- and cryogenic-safe nano-positioner, which sits at the focus of a 50x objective. To determine the position of the graphene device, we use a Raman map of the graphene's 2D phonon peak, as well as the Raman signal from the Si substrate (see **Supplemental Information**). After the Raman map, we perform Procedure (1) at $V_{BG}$ = 0 V to "clean" the sample of any charge doping inhomogeneity created by the small laser spots during the mapping step. Then, a fixed backgate voltage, $V_{D-L}$, was applied while a portion of the graphene heterostructure was illuminated with a 633 nm laser focused through the same 50x objective. Specifically, the sample was rastered relative to the fixed laser position over a locally defined region for 10 minutes for each doping level (i.e. backgate voltage). At best, the laser spot size is diffraction limited, with a Gaussian intensity profile of approximately 1 micron in diameter. After the light source was extinguished, the backgate voltage was set to zero to conclude the doping procedure. Procedure (2) is shown schematically in **Figure 2A**.

**Transport Measurements**

Magnetotransport measurements were performed at 1.6 K with magnetic fields between 0 T and 9 T applied out-of-plane to verify functionality and quantify the doping properties of the device. The device was only illuminated during the doping processes. During transport measurements, all illumination sources (white light illuminator or laser) were off. Traditional lock-in amplifier techniques were used with currents ranging from 5 nA to 100 nA at 19 Hz.[32,38] The estimated mobility was $\sim$ 100,000 V cm$^{-2}$ s$^{-1}$ at 1.6 K for a carrier density of 1x10$^{12}$ cm$^{-2}$ determined from a conventional Drude model.[39] The longitudinal resistance (R$_{XX}$) and Hall conductance (G$_{XY}$) were measured as a function of applied backgate voltage (V$_{BG}$), global (V$_{D-G}$) and local (V$_{D-L}$) doping voltages, and applied magnetic field, B. After each transport measurement, regardless of whether the sample was locally or globally doped, the sample was illuminated with white light overnight with an applied V$_{D-G}$ = 0 V. This acted as a "reset" for the sample to erase previous doping configurations and restore the device to its original response. For all of the doping experiments presented in the main text, we used negative V$_{D-G}$ or V$_{D-L}$ values. However, this doping technique also works with positive V$_{D-G}$ or V$_{D-L}$ (see **Supplemental Information**).

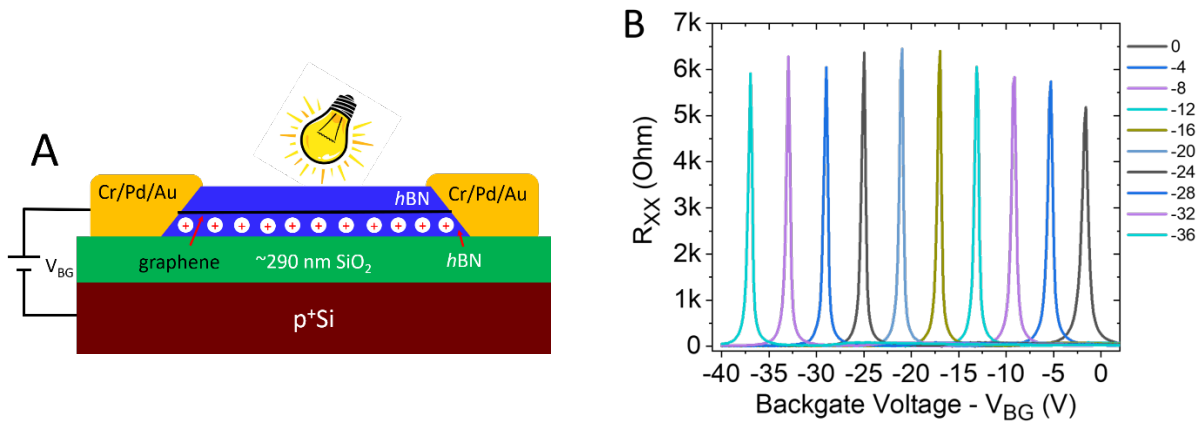

**Figure 1:** **A**: Schematic cross-section of the device and the biasing scheme during the global optical doping process (Procedure (1)). **B**: Channel resistance *vs* backgate voltage at different global doping voltages ranging from V$_{D-G}$ = 0 V to V$_{D-G}$ = -36 V in 4 V steps. The position of the Dirac peaks at each V$_{D-G}$ could be accurately determined to be V$_{Dirac\_x}$ = V$_{Dirac\_0}$ + V$_{D-G}$ where V$_{Dirac\_0}$ $\approx$ - 1.7 V.

**Results and Analysis**

**Basic Transport Results**

First, we demonstrate and analyze the properties of a globally doped device. These results are shown in **Figure 1B**, where the channel's resistance, $R_{XX}$, is plotted *vs* $V_{BG}$ from a series of global doping procedures. In this sequence, Procedure (1) was repeated at different values of $V_{D-G}$ from 0 V to – 36 V at 4 V steps. The device's response at $V_{D-G} = 0$ V is the baseline value and is used to compare all other doping levels. The baseline graphene Dirac point is $V_{Dirac\_0} \approx$ -1.7 V, and this value stayed constant over the course of the experiment duration of $\sim$ 3 months. The position of the Dirac point after illumination while a given global voltage, $V_{D-G} = x$, is simultaneously applied is $V_{Dirac\_x} = V_{Dirac\_0} + V_{D-G}$, is determined by the doping level.[25]

It is clear from these data that the ultimate doping level can be set solely by the $V_{D-G}$. We determine the n-doping level in the sample for each $V_{D-G}$ which ranged from $\sim5x10^{10}/cm^2$ at $V_{D-G}$ = 0 to $\sim2x10^{12}/cm^2$ at $V_{D-G}$ = –36 V. (See **Supplemental Information** for calculation details and the doping values for each $V_{D-G}$.) The remarkable series of peaks in **Figure 1B** illustrates the precise doping control that can be achieved for each $V_{D-G}$. This determination of the $V_{Dirac\_x}$ (proportional to doping level), allows us to reproducibly tune the global doping level in the device with very high precision and dynamic range.

While an arbitrary area can be doped with the laser-based local doping approach described in **Procedure (2)**, we chose the simple geometry shown in the shaded area of the schematic in **Figure 2A** to demonstrate this method. As with global doping, the doping level of the locally illuminated area is determined by the value of $V_{D-L}$, while the rest of the sample (the unilluminated global region) will have no doping change. In **Figure 2B**, the green curve is the channel resistance, $R_{xx}$, as a function of backgate voltage, $V_{BG}$, after the local region of the sample was subjected to photodoping with $V_{D-L} = – 18$ V. A clear indication of successful spatial doping is the emergence of two Dirac peaks in the measurement. One peak, $V_{BG} \sim V_{Dirac\_-18}$, is the response of the local region and the other is the Dirac peak at $V_{BG} \sim V_{Dirac\_0}$ representing the response of the global region of the sample. To systematically change the doping level in the local region, we repeated **Procedure (2)** at different $V_{D-L}$. The overall doping configuration of the sample with respect to the backgate voltage is then changed for each local doping level. In the example shown in **Figure 2B** with $V_{D-L} = –18$ V, from left to right, the yellow, red and blue shaded areas correspond to $p^+$-p-$p^+$, p-n-p, and n-$n^+$-n, sample doping configurations respectively. For simplicity, we hereafter abbreviate these junctions as p$^+$-p-p$^+$=*PpP*, p-n-p=*pnp* , and n-n$^+$-n=*nNn*. Changing V$_{\text{D-L}}$ used in **Procedure (2)** changed the backgate voltage range where the device remained in each of these 3 different doping configurations.  For example, with fixed backgate voltage ranging from -40 V to 40 V and with V$_{\text{D-L}}$ decreasing to a more negative value, the backgate voltage range where the sample stayed in the *pnp* doping configuration increased, and the backgate voltage range when the sample stayed in the *PpP* doping configuration decreased.

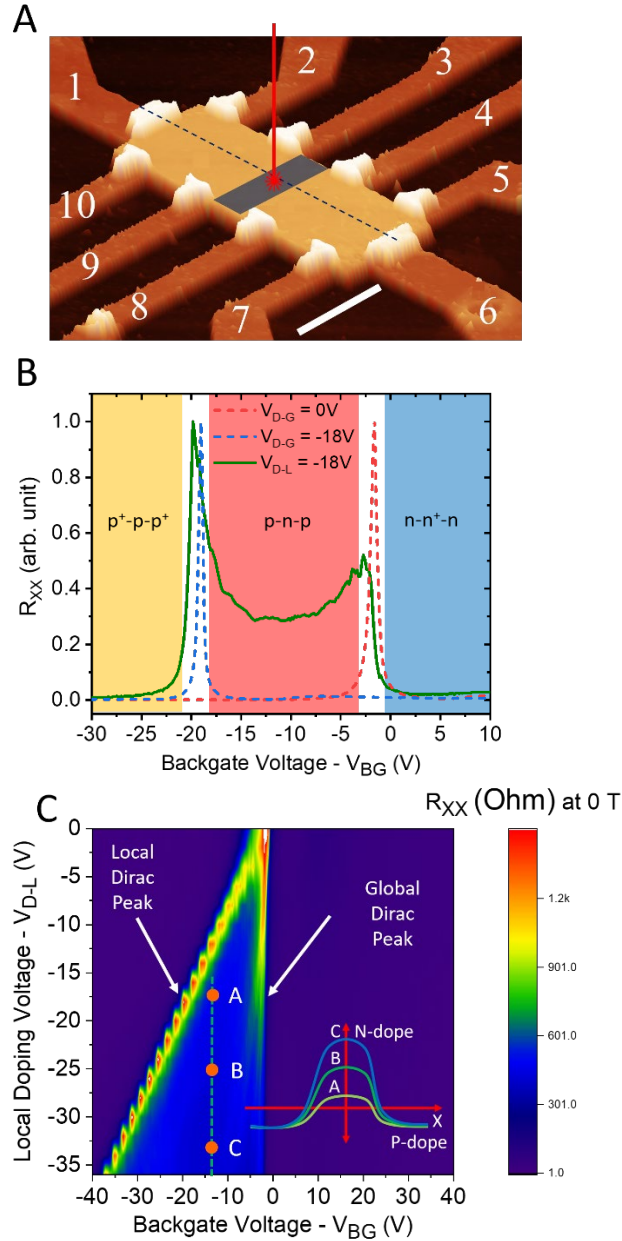

**Figure 2: A**: AFM image of the device after contact deposition and lift-off, scale bar is 5 μm. The illustrated, shaded area in the middle represents the local doping region of the device created by the combination of laser illumination and backgate bias (doping Procedure (2)). The rest of the sample is the global region. The dark green dashed line marks the position of the cross-sectional cut of the device that is shown schematically in **Figure 1A**. **B**: Normalized resistance of the device as function of backgate voltage when it was globally doped at $V_{D-G}$ of 0 V and -18 V respectively, red and blue dashed curves, and locally doped at $V_{D-L}$ = -18 V, green curve. Yellow, red, and blue shaded areas on the plot indicating three different regimes of spatial doping configuration with respect to the backgate voltage: p+pp+, pnp, and nn+n. **C**: two-dimensional map of channel resistance as function of backgate voltages, $V_{BG}$, and local doping voltages, $V_{D-L}$. Arrows mark the positions of the global Dirac peak and local Dirac peak, respectively. Inset shows schematic of the *pnp* junction electrostatic profile with respect to the different n-doping level in the local region (different $V_{D-L}$) at fix backgate bias, point A, B, and C on the main figure.

**Figure 2C** shows the 2D map of channel resistance, $R_{XX}$, *vs* backgate voltage, $V_{BG}$, and local doping voltage, $V_{D-L}$. The positions of the Dirac peak are indicated by white arrows with the global Dirac peak on the right and local Dirac peak on the left. The location of the local Dirac peak evolved linearly with changing $V_{D-L}$, indicating consistent and accurate doping of the local region over the whole $V_{D-L}$ range. The global Dirac peak remained unchanged, indicating that the doping level of global portion device was not affected by the local doping procedure. Inset shows schematic of the doping distribution along the sample in *pnp* configuration with respect to different n-doping level in the local region (different $V_{D-L}$) at fix backgate bias, point A, B, and C on the Figure 2C.

**Quantum Hall Effect**

We performed quantum Hall transport measurements on our spatially-doped junctions after each doping experiment to characterize the doping interfaces. Due to the unique semi-metal properties of graphene that arise from its linear Dirac band structure, pn-junctions (pnJs) in graphene do not exhibit the unidirectional current-blocking behavior at zero magnetic fields associated with traditional pn-diodes. When a strong perpendicular magnetic field is applied to a graphene pn-device, bulk electrons and holes in the p- and n-regions of the device are strongly localized by the magnetic field. The device transport properties are governed by the formation of chiral Landau level (LL) edge-state channels circulating along the boundary and the pnJ interfaces.[2,38,40] The LL edge channels circulate in opposite directions for p- and n-regions, resulting in edge states of opposite charge traveling in the same direction, alongside each other, at the electrostatic pnJ interfaces. In the quantum Hall regime, the Hall resistance, $R_{XY}$ is quantized to values of $(1/\nu)(e^2/h)$ where $\nu$ is an integer, e is the fundamental electron charge, and h is Plank's constant. For graphene, typically, $\nu = \pm2, \pm6, \pm10, \ldots$[2] giving rise to $R_{XY}$ plateaus when $\nu_G$ and $\nu_L$, the filling factors in the global and local regions, respectively, are near these integer values. $\nu_G$ and $\nu_L$ are determined by the carrier concentration in the global and local regions, respectively, divided by the density of carriers per LL which depends on the applied magnetic field. When these characteristic quantum Hall resistance plateaus are observed the longitudinal resistance $R_{XX}$ is simultaneously 0 Ω. Previous quantum Hall measurements of graphene pnJ devices [38,40] showed that LL interactions at the pnJ interfaces can be used to characterize the electrostatic profile of the pnJ. We used similar measurement and analysis methods here to characterize the pnJs created with the spatially resolved photodoping technique.

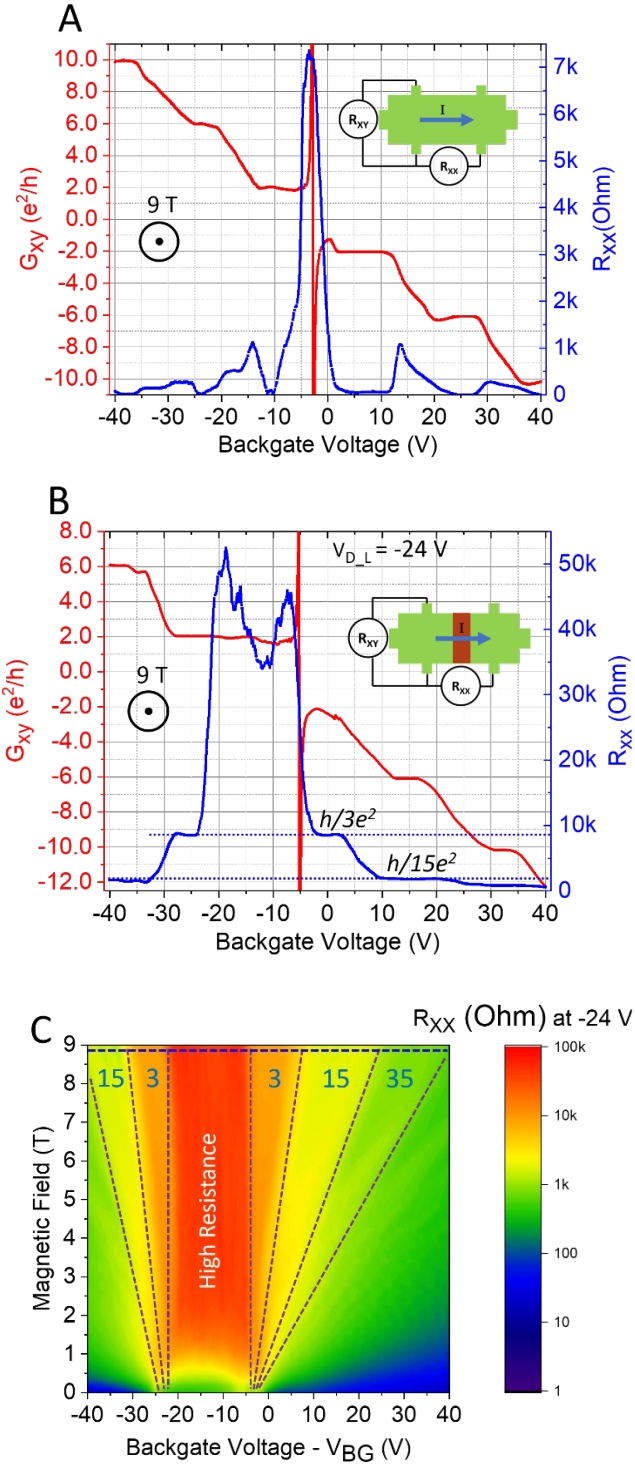

**Figure 3**: Quantum Hall response of the device when it is **A)** uniformly doped (**Procedure 1**) at $V_{D-G} = 0$V, and **B)** when it is spatially doped (**Procedure 2**) at $V_{D-L} = -24$ V. Red dashed lines in B) show the position of quantum Hall plateaus in the longitudinal resistance. **C)** Landau fan diagram of the longitudinal conductance of the sample after it was spatially doped with **Procedure 2** and with $V_{D\_L} = -24$ V. Dashed lines are guide for the eyes and mark the transition between different quantized resistance plateaus. The numbers in between the lines show the filling factor of the respective plateaus. Horizontal dashed line is the profile cut of the data shown in **B**.

 Figure 3A illustrates the typical quantum Hall effect in the device when it is uniformly doped at $V_{D-G} = 0$ V (so that the position of the Dirac point is at $V_{Dirac\_0}$) following **Procedure 1**, as a function of backgate voltage, $V_{BG}$. The inset in **Figure 3A** shows a schematic of the measurement configuration. The blue curve is the longitudinal resistance, $R_{XX}$, with values shown on the right y-axis. The red curve is the Hall conductance, $G_{XY}$, in units of $e^2/h$, with values shown on the left y-axis. The device shows a clear quantum Hall response at 9 T where we observe the quantized Hall conductance at values of $\pm\ 2e^2/h, \pm\ 6\ e^2/h,$ and $\pm\ 10\ e^2/h$. We observe positive values for electrons and negative values for holes, consistent with the measurement configuration and direction of the magnetic field. $R_{XX}$ is zero at the quantized nodes in the Hall conductance. This quantum Hall response agrees with previous measurement of similar heterostructures [41,42]. We used these measurements to determine the filling factors of the device that will be used in the analytical model detailed later in this paper.

 **Figure 3B** shows the quantum Hall response of the sample, at 9 T, after **Procedure 2** was used to create a local doping configuration with $V_{D-L} = -24$ V. The position of the locally doped region is indicated by the brown shaded area in the measurement schematic inset of **Figure 3B**. The two horizontal dashed lines in **Figure 3B** highlight two resistance plateaus in $R_{XX}$. For the region of the device outside of the locally doped region, the doping level remains unchanged at $V_{BG} = 0$V. Besides the regular quantized conduction plateaus in the Hall conductance, the dominant feature of this measurement is the appearance of unconventional quantized resistance values of $h/3e^2$, $h/15e^2$, and $\approx h/35e^2$ in $R_{XX}$. These quantized resistance steps in $R_{XX}$ only occur when the sample is in the *PpP* and *nNn* doping configurations which arise when $V_{BG} < -24$ V and $V_{BG} > -1.7$ V, respectively. A model for the LL edge state transport schematic of these two cases is shown in **Figure 4A** and **Figure 4B**. Detailed analysis of the plateau values will be presented later. However, from the transport schematic (shown in **Figure 4A** and **Figure 4B**), the quantization of the longitudinal resistance is only possible when the device has a complete and clean doping interface across the entire width of the sample. These clean interfaces enable the transparent transmission of low filling factor LL edge states across between doping regions while completely blocking the higher filling factor LL edge states from being transmitted. These quantized steps in $R_{XX}$ are therefore evidence that the pnJs were completely and cleanly formed across the width of the device and hence play a crucial role in the longitudinal resistance conductance's quantization.

To further elucidate the magnetotransport properties of our device in the spatial doped configuration, we performed full Landau fan diagram measurements. In these measurements, the resistance of the device was measured as a function of $V_{BG}$ and magnetic field. **Figure 3C** shows the Landau fan diagram at $V_{D-L} = -24$ V. The horizontal dashed line near the top is the profile cut of the data shown in **Figure 3B**. The angled dashed lines are a guide for the eye, separating regions of the resistance steps that have different quantized values. The numbers between the lines are the respective filling factors of each region. Noticeably, the red area in the middle of the figure, in the range of -1.7 V < $V_{BG}$ < -24 V, is a high resistance (insulating) state of the sample where the doping configuration is *PnP*. This insulating behavior persisted as along as the sample was in the quantum Hall regime down to B ~1 T. At lower field, when the sample was not in the quantum Hall regime, the insulating state disappeared. In the high resistance state, the resistance value of the device was as high as ~100 kOhm and did not show any well-defined resistance quantization indicative of quantum Hall conduction. This "insulating" state is not common and has not been observed in previous transport measurements of graphene pnJ devices in high magnetic fields.[43–46] To the left of the high resistance region of the magnetoresistance plot, the sample doping is *PpP*, and to the right of the high resistance area, the sample is doped *nNn*.

In quantum Hall devices with a differently doped region that extends across the entire width of the sample such as our devices here, the resistance measured across the "barrier" region can depend strongly on the details of the coupling or equilibration between the various LL edge states. In previous experiments using graphene pnJ devices created by conventional lithographically patterned gate structures, two distinctive LL edge state equilibration regimes were realized: (i) full equilibration of LL edge states at the pnJ interface,[40] (ii) and lowest LL edge state equilibration at the pnJ interface. [38] Theoretical calculations[47] indicate electrostatically sharp junctions where the transition between the differently doped regions is spatially abrupt enable full equilibration while a more gradual spatial transition in an electrostatically graded junction promotes lowest LL equilibration. The quantum Hall transport of the pnJs in our spatially photodoped devices is therefore an important tool to identify the electrostatic profile of the transition between locally doped regions and will be discussed further below.

**Landauer-Buttiker Edge State Model**

To better understand the quantum Hall response of our device, we compare our experimental results with calculations based upon the well-known Landauer-Büttiker edge state formalism. [31,48] In order to carry out this comparison, we calculated the expected quantized resistance values with respect to different doping configurations and quantized filling factors.

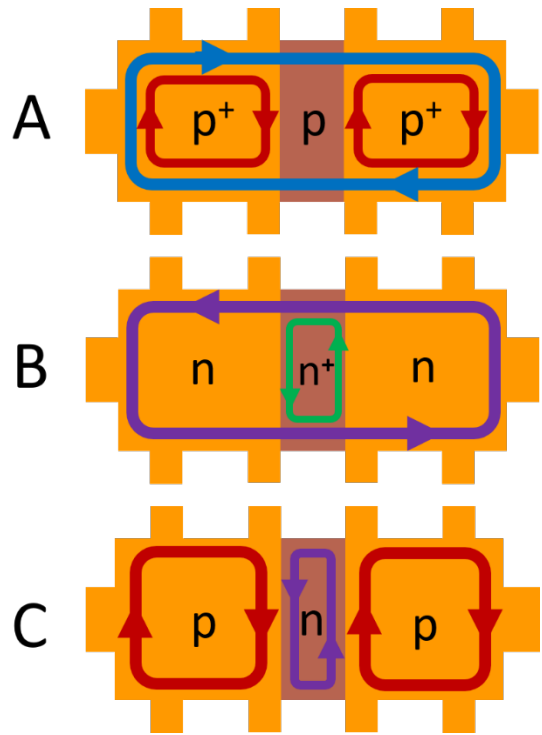

**Figure 4:** Visualization of the Landau level edge states transport in the device in the quantum Hall regime with the local doping configuration at different backgate voltages. From left to right of the backgate voltage range, three doping configurations were realized: **A**: *PpP*, **B**: *nNn*, and **C**: *pnp*.

The edge state current flow in the quantum Hall regime as modeled in Landauer-Büttiker formalism is schematically shown in **Figure 4** for the various electrostatic configurations where the local and global regions of the graphene are doped at different levels and polarities. The doping type and qualitive levels can be determined by the position of the Dirac peak in the global/local region relative to the position of the Fermi level which is set for by the back gate. For example, in the left side the back-gate range of the $R_{XX}$ resistance maps identified in yellow in **Figure 2B** and shown in **Figures 3B, C** the device is in a unipolar *PpP* doping configuration. The device is p-type in both the global and local regions with the doping level (filling factor) higher, *P*, in the global region relative to the doping level (filling factor) in the local region, *p*. In this gating range, $\nu_G > \nu_L$. The current flow can be envisioned in this model as follows. When the chiral edge current is incident on the locally p-doped region, the lower index LL's corresponding to the more lightly doped local region flow unimpeded into the locally doped region as schematically illustrated in **Figure 4A**; however, the higher LL's cannot transmit into the locally doped region as no states are available. These states will scatter at the interface between the local and global regions and cross the sample to the other edge where they reverse direction and are effectively backscattered. On the far side of the local region, all the current in the populated edge states can continue into the more heavily doped global region and continue to the other current contact.

For this unipolar *PpP* doping configuration values the observed quantized $R_{XX}$ resistance plateaus can be calculated within the Landauer-Büttiker edge state formalism (as fully described in the **Supplemental Information**).

$$R_{xx} = \frac{h}{e^2} \frac{\nu_G - \nu_L}{\nu_G \nu_L} \qquad (1)$$

Similarly, for the unipolar *nNn* doping configuration, corresponding to the right side of the backgate range in a magnetoresistance plot or map (indicated in blue in **Figure 2B**), the device has n-type doping in both the global and local regions. The doping level (filling factor) in the global region is lower than the doping level (filling factor) in the local region such that $|\nu_L| > |\nu_G|$. In this configuration all the chiral edge current incident on the more heavily n-doped local region will transmit into this locally doped region as schematically illustrated in **Figure 4B**. It has been experimentally observed that in the integer quantum Hall regime, current rapidly equilibrates across all available LL edge states in graphene devices along mechanically etched edges [49,50] Therefore, once in the locally doped region, the current will rapidly redistribute across the available edge channels. When these chiral edge states reach the far interface between the local and global regions, the higher LL's cannot transmit out of the locally doped region as no states are available in the global region. Just as described above, these states will scatter at the interface between the local and global regions and cross the sample to the other edge where they reverse direction within the local region (**Figure 4B**). The quantized $R_{XX}$ resistance plateaus calculated in the quantum Hall regime within the Landauer-Büttiker edge state formalism for this unipolar *nNn* doping are,

$$R_{xx} = \frac{h}{e^2} \frac{|\nu_L| - |\nu_G|}{|\nu_G \nu_L|} \qquad (2).$$

See **Supplemental Information** for the derivation of equations (1) and (2) and tabulated values as function of filling factor for each of these unipolar doping conditions.

The calculated quantized $R_{XX}$ values for both the unipolar PpP and nNn doping conditions match the experimental values reported in the 2D resistance maps shown in **Figure 3C.** The observation of these "unconventional", quantized values of $R_{XX}$ and the quantitative agreement with a simple edge-state model provides strong evidence of the promising capabilities of this optical doping method. The technique has both excellent electronic and spatial control of the doping properties of the 2D graphene device. The agreement between the model and measured data is only possible when the locally doped region completely crosses the full width of the device and there are good interfaces between the local and globally doped regions.

As was mentioned earlier, the quantized resistance values measured across the "barrier" region can depend strongly on the details of the equilibration between the various LL edge states as they travel along the interface between the two differently doped regions. However, for the PpP and nNn unipolar cases just discussed, where resistance plateaus are observed in $R_{XX}$, the mechanism of quantized resistance plateau formation can be explained by using the Landauer-Büttiker edge state formalism with no assumptions regarding the interaction of the edge-states at the interfaces between the two doping regions. The LL edge states that can be accommodated in the lower carrier density region circulate the entire device passing through both the global and local regions. Additional edge states arise in the higher carrier concentration regions that do not cross the doping interfaces and circulate only in the regions of higher filling factors.

In the case of a bipolar regime, or *pnp* doping, corresponding to the middle part of the back gate bias region (shown in **Figure 2B**, red shaded area), the LL edge-states in the p- and n-regions circulate in opposite directions in the global and local doping regions. Therefore, they travel alongside each other in the same direction along the left and right interfaces between the global and local regions, as demonstrated schematically in **Figure 4C**. It was previously confirmed that [38] in such a case, depending on the electrostatic profile of the pnJs, there could be at least two possible ways the p- and n-LL edge states could interact with each other across the pnJ interfaces.

Quantized resistance values for the longitudinal resistance can be calculated according to Landauer-Büttiker edge-state formalism for each of these two cases. In one case, the current equilibrates among all the edge channels on both sides of the interface between the global and local regions, and, in the other case, only the lowest LL exchanges current across the interface. See **Supplemental Information** for a detailed explanation of the models for the *pnp* configuration for both equilibration cases.

For **Case (1):** The p- and n-LL edge states completely equilibrate across the pnJ interfaces allowing all LL edge states to participate in conducting current across the device from source to drain. The longitudinal resistance in the integer quantum Hall regime with full equilibration, p-doped in the global region with filling factor of $\nu_G$ and n-doped in local region with filling factor of $\nu_L$ is given by:

$$R_{xx} = \frac{h}{e^2} \frac{|\nu_L| + |\nu_G|}{|\nu_G \nu_L|} \qquad (3)$$

For **Case (2):** Only the lowest LL edge states are equilibrated across the pnJ interfaces to conduct current across the regions. All the higher LL edge states are not equilibrated with the lowest LL edge state and backscatter across the sample at the pnJ interfaces. The longitudinal resistance with partial equilibration, p-doped in the global region with filling factor of $\nu_G$ and n-doped in local region with filling factor of $\nu_L$, lowest LL edge state with filling factor of $\nu_0 = 2$ is given by:

$$R_{xx} = \frac{h}{e^2} \frac{3|\nu_G| - |\nu_0|}{|\nu_G \nu_0|} \qquad (4)$$

The measured resistance values of the device when it is in the *pnp* doping configuration in the quantum Hall regime, shown in **Figure 3B** and **C**, do not match the values predicted by either of these two LL edge state interaction models. Quantized conduction is not observed, and the resistance values are much higher than the expected values according to the models for any filling factor. This high resistance is indicating that the central n-doped region is acting to block the current in this device configuration.

In previous experiments,[38,40,43] it was shown that the electrostatic profile of the pnJs plays a crucial role in determining the interaction between n- and p-LL edge states at the pnJ interfaces. In physically abrupt or sharp pnJ interfaces the p- and n-LL edge states in are close proximity to each other as they travel alongside each other at the pnJ interfaces. This proximity enables the scattering of carriers between the edge states and the equilibration of the chemical potential all of the LL's edge states participate in conducting current across the pnJs. Alternatively, in a more graded junction, where the n- and p-LL's level edge states are physically further apart as they travel along the doping interfaces, only the energetically lowest n- and p-LL's edge states interact. This physical separation results in a partial equilibration of the LL's edge states across the junction. Due to the nature of the optical doping technique, we expect it will create graded pnJ junctions. In the best-case scenario, the Gaussian intensity profile of the laser beam creates a doping region with edges that diminish as $1/e^2$, over approximately 1 um.  This optical gradient implies that the pnJs formed are gradually graded compared to conventional gate-defined pnJs. See **Supplemental Information** Figure S7 for schematic demonstration of the electrostatic profile of the pnJs created by this doping technique in relative comparison with pnJs created by other techniques shown in Ref. 38 and Ref. 40 respectively.

The spatial photodoping profile of an $h$BN/Gr/$h$BN heterostructure was previously inferred by gradually extending the local doping region between two voltage probes of the device while monitoring the evolution of the Dirac peaks.[26] This method, while given an indication of successful spatial doping, does not provide detailed information on the quality of doping profile of the whole device nor how complete the doping boundary is across the width of the sample. Also, the estimation of the spatial electrostatic profile of the doping interface is limited by the physical dimensions and geometry of the voltage probes as well as the alignment of the laser with respect to these voltage probes.

The quantum Hall measurement indicates that there is little or no interaction between the LL edge states between the p-region (global region) and the n-region (local region) and there is no complete transmission of any LLs, not even the energetically lowest. The device is in a current-blocking state at this doping configuration.

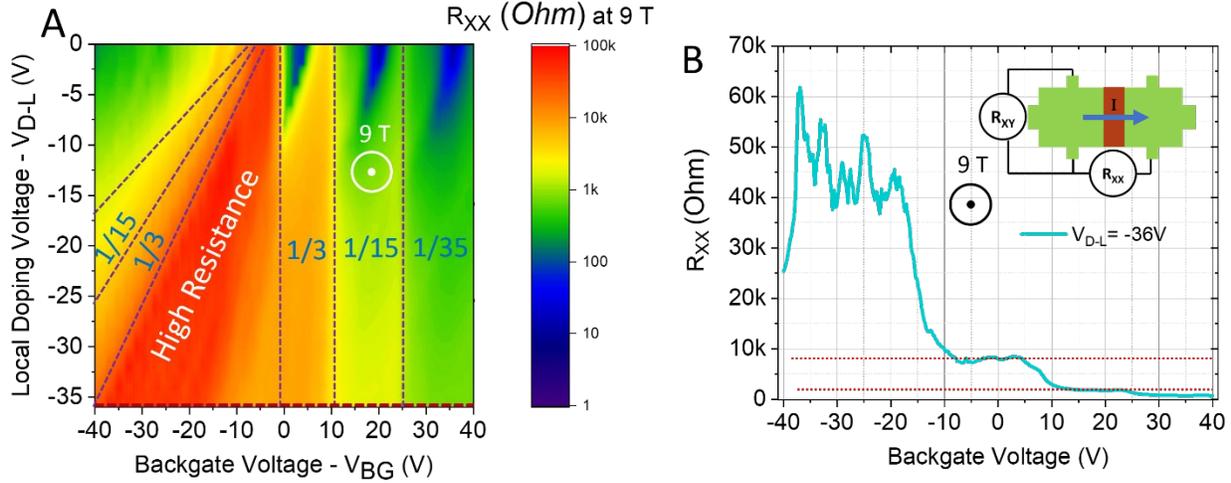

**Figure 5: A:** 2D map of $R_{XX}$ *vs* back gate voltage, $V_{BG}$ and local doping voltages, $V_{D-L}$. Purple dashed lines are guide for the eyes and mark the transition between different resistance plateaus. The brown line near the x axis corresponds to the profile cut shown in **B**. **B:** Longitudinal resistance, $R_{XX}$, of the sample at high magnetic field, 9T, in the spatial doping configuration with $V_{D-L}$ = -36 V.

To further investigate the highly resistive state of the sample in the *pnp* doping configuration in the quantum Hall regime we performed extensive doping modification of the local region and measured the respective quantum Hall response after each doping change. Shown in **Figure 5A** is the quantum Hall resistance map of $R_{XX}$ *vs* $V_{BG}$ and local doping gate, $V_{D-L}$. The magnetic field was fixed at 9 T. The purple dashed lines are a guide for the eyes, separating regions of the resistance steps that have different quantized resistance values. The red area in the middle of the figure is the high resistance state of the sample where the doping configuration is *pnp*. To the left of the high resistance area, the sample was in the *PpP* doping configuration, and to the right of the high resistance area the sample is in the *nNn* regime. By increasing the n-doping density in the local region, the backgate bias range in which the sample stays in the *pnp* doping configuration increases and expands towards the negative back gate bias. Regardless of the doping density level in the local region, we observed well-quantized resistance plateaus when the sample is outside of the *pnp* range and in the *PpP* and *nNn* configurations. This quantized behavior confirms that clean and complete interfaces are formed between the local and global region of the sample, and are further proof of the repeatability and stability of this doping technique over a large doping range. No quantized resistance plateaus in $R_{XX}$ were observed when the sample was in the *pnp* doping configuration regardless of the doping density in the local region. $R_{XX}$ consistently stays in the high resistance, current blocking, state throughout the doping range when sample is in *pnp* spatial doping configuration. This observation reinforces our initial assessment that the photodoping technique has created a gradually graded pnJ profile in our device.

The unique electrostatic profile of the pnJ interfaces created by the photodoping gives rise to the novel observed transport behavior that was just described above. Despite its wonderful electrical properties, graphene lacks a bandgap and associated gate-toggled high and low resistance states. Therefore, practical applications in digital electronic architectures have not been forthcoming. Our unique junction profile combined with low quantization B-field requirement could create a high-performance graphene switching device that has a good ON/OFF ratio. In addition, this gradually graded pnJ profile, when also demonstrated in other 2D TMD semiconductors interfaces with *h*BN, could reveal many other possible applications, such as pnJ light emitting diodes or solar cells.

## Conclusion

The ability to accurately control and quantitively measure the carrier concentration in layered 2D semiconductors is critical for practical applications of this novel class of materials. We demonstrate that by optically activating/deactivating the chargeable trap states in *h*BN, we can use them as remote electrostatic remote dopants to enable control and reversible dope graphene in a *h*BN/Gr heterostructures. More importantly, using a spatially resolved light source, these trap states could be activated/deactivated with accurate spatial distribution allowing us to create modulational doping of graphene and realize pnJ devices. Uniquely, we used low temperature, *in-situ* quantum Hall transport measurement to characterize the pnJs at various doping density and magnetic field strengths. Our magnetotransport measurements 1) confirmed the high quality of the optically doped interfaces; 2) proved the grading of the doping profile and thereby the separation of the p and n-LL; and 3) demonstrated that properly engineered doping configurations can turn gapless graphene into a functioning ON/OFF switch at moderate magnetic fields.

## Competing Interests

The authors declare no competing financial interests.

## Acknowledgements

The authors from LPS gratefully acknowledge assistance from the LPS support staff including G. Latini, J. Wood, R. Brun, P. Davis and D. Crouse.